# Accurate DFT Simulation of Complex Functional Materials: Synergistic Enhancements Achieved by SCAN meta-GGA

Da Ke[1], Jianwei Sun[2,*], and Yubo Zhang[1,*]

[1]*Minjiang Collaborative Center for Theoretical Physics, College of Physics and Electronic Information Engineering, Minjiang University, Fuzhou 350108, China*

[2]*Department of Physics and Engineering Physics, Tulane University, New Orleans, Louisiana 70118, USA*

Corresponding email address: yubo.drzhang@mju.edu.cn & jsun@tulane.edu

Complex functional materials are characterized by intricate and competing bond orders, making them an excellent platform for evaluating the newly developed *Strongly Constrained and Appropriately Normed* (SCAN) density functional. In this study, we explore the effectiveness of SCAN in simulating the electronic properties of displacive ferroelectrics ($BaTiO_3$ and $PbTiO_3$) and magnetoelectric multiferroics ($BiFeO_3$ and $YMnO_3$), which encompass a broad spectrum of bonding characteristics. Due to a significant reduction in self-interaction error, SCAN manifests its improvements over the Perdew-Burke-Ernzerhof (PBE) method in three aspects: SCAN produces more compact orbitals, predicts more accurate ionicity, and better captures orbital anisotropy. Particularly, these synergistic enhancements lead to notable phenomena in calculating the bandgap of $YMnO_3$: while the PBE+$U$ simulation may suggest a strong correlation appearance attributed to high Hubbard-like $U$ values (~5 eV), the value is dramatically lower (~1 eV) in the SCAN+$U$ method. Furthermore, we provide an intuitive analysis of SCAN's operational principles by examining the complex electron densities involved. These insights are theoretically intriguing and have practical implications, potentially encouraging wider adoption of SCAN in the computational modeling of complex functional materials.

## 1. INTRODUCTION

Complex functional materials are designed to fulfill specific applications, such as ferroelectrics, multiferroics, piezoelectrics, thermoelectrics, magnetoresistive materials, shape memory alloys, high-temperature superconductors, and quantum dots [1]. These materials often require intricate chemical bonds at the atomic scale or complicated structural integration on larger length scales to realize fine tunability in response to external stimuli. For example, ferroelectric materials have a reversible spontaneous electric polarization due to their non-centrosymmetric atomic arrangement [2]. Multiferroics often combine electrical and magnetic properties through more complex crystal structures [3]. High-efficiency thermoelectric materials frequently depend on sophisticated structural engineering to independently optimize electrical and thermal properties, which typically present conflicting material attributes [4]. For high-temperature cuprate superconductors [5], the complexity arises from the geometry and the presence of many intertwined electronic phases associated with the strongly correlated electrons.

While Kohn-Sham density-functional theory (DFT) [6,7] is widely used in simulating materials, complex functional materials still present significant challenges to conventional DFT functionals. The difficulties are evident, for example, in displacive-type ferroelectric and magnetoelectric multiferroic materials. For instance, $BaTiO_3$ loses its centrosymmetry upon cooling, and the strength of this distortion arises from anisotropic bonding around Ti ions [2], where each bond displays a distinct combination of covalent and ionic characteristics [8]. Achieving a reliable balance between these two types of electronic interactions poses significant challenges for DFT functionals. In the case of multiferroic $BiFeO_3$, complexity increases with the addition of atomic magnetization, which originates from the localization of *d*-orbitals. Despite their localization, these *d*-orbitals inevitably hybridize with the delocalized O-2*p* electrons, a critical process for achieving magnetoelectric coupling. Managing this delicate balance between electron localization and orbital hybridization continues to be a challenge in DFT calculations.

The most widely adopted DFT functionals have been the *local density approximation* (LDA) [9,10] and the *generalized gradient approximation* (GGA) as exemplified by the *Perdew-Burke-Ernzerhof* functional (PBE) [11]. To address the notable *self-interaction error* associated with the *d*-orbitals in local/semilocal functionals, a Hubbard-like $U$ correction is commonly applied, and the $U$ values are often empirically determined to replicate expected properties. A frequent practice is employing DFT+$U$ to achieve a positive bandgap in insulators. We will show that such empirically fitted $U$ values can be excessively high, leading to significant distortions in the band structure when compared to experimental results. This may adversely affect other properties of the functional materials. For instance, in calculating the bandgap of YMnO$_3$, the PBE+$U$ simulation suggests a strong correlation appearance due to high $U$ values (5~7 eV); moreover, the high $U$ nearly suppresses the crucial *pd* hybridization that is essential for magnetoelectric coupling.

A recent advancement is the *strongly constrained and appropriately normed* (SCAN) meta-GGA [12], which has attracted considerable interest in the computational materials science community and has been adopted in the latest Materials Project [13]. The SCAN functional incorporates kinetic energy density into its formulation, enhancing its ability to distinguish between different types of chemical bonds [14] and reducing the *self-interaction error* [15]. Previously, we have demonstrated how SCAN can surpass PBE in accurately describing some simple transition-metal compounds. For example, TiO$_2$ features several energetically near-degenerate phases; SCAN significantly outperforms other popular functionals in differentiating these phases [16]. FeO often challenges local/semilocal functionals, which typically fail to open the bandgap unless augmented with the Hubbard-like $U$ correction. However, FeO demonstrates a successful bandgap opening in our SCAN calculations without $U$ [17], which has sparked renewed discussions about DFT's effectiveness in describing the correlated electrons [18-20]. SCAN also demonstrates superior performance in calculating the electronic properties of *d*-orbital cuprates [21-23] and *f*-orbital SmB$_6$ [24], as well as the ferroelectric properties [25].

In this work, we underscore three synergistic enhancements the SCAN functional provides in modeling complex materials: it produces more compact orbitals, predicts more accurate ionicity, and better captures orbital anisotropy than PBE. This synergy leads to a more accurate description of ferroelectric and multiferroic materials, including identifying an alternative gapping mechanism in YMnO$_3$. Furthermore, our analysis delves into the connection between SCAN's enhanced predictive capabilities and its detailed representation of electron density distributions. This intuitive insight may potentially guide the computational materials community in using SCAN to model complex materials and help develop new functionals to address density-driven errors [26].

## 2. COMPUTATIONAL DETAILS

Simulations were conducted using the Vienna Ab-initio Simulation Package (VASP) [27], employing the projector-augmented wave method [28,29]. For comparative analysis, we utilized the LDA [6,7], PBE GGA [11], and SCAN meta-GGA [12,14]. Additionally, we applied the HSE06 hybrid functionals [30,31] to specific systems. For the elements Ti, Mn, and Fe, we included the semicore *p*-states as valence states; for Pb and Bi, we treated the semicore *d*-states as valence states. We employed an 8 × 8 × 8 Γ-centered K-mesh for the five-atom cells of BaTiO$_3$ and PbTiO$_3$, and a 4 × 4 × 2 K-mesh for the thirty-atom cells of BiFeO$_3$ and YMnO$_3$. Regarding magnetic configurations, we set G-type antiferromagnetism to BiFeO$_3$ and A-type antiferromagnetism to YMnO$_3$. Lastly, we omitted spin-orbit coupling effects from all systems. The valence state was derived from the Bader charge analysis [32].

# 3. RESULTS

## 3.1. Bonding anisotropy and ionicity

Displacive-type ferroelectric BaTiO$_3$ and PbTiO$_3$ lose centrosymmetry upon cooling. The distortion's strength stems from anisotropic bonding around Ti ions: While the elongated Ti-O bond exhibits increased ionic attributes, the opposite shortened bond has enhanced covalent properties. The off-center atomic displacement leads to a tetragonal deformation characterized by $\eta = c/a > 1$, where $a$ and $c$ are the lattice constants. The calculated $\eta$ sensitively depends on the DFT functionals (Figure 8b): whereas $\eta$ is relatively well calculated by LDA, it is strongly overestimated by PBE, known as the *super-tetragonality* problem [33]. For instance, the PBE functional overestimates $\eta = 1.054$ for BaTiO$_3$, compared to the experimental low-temperature value of ~1.03 [25]. By contrast, recent SCAN calculations have predicted more accurate results, with $\eta = 1.029$.

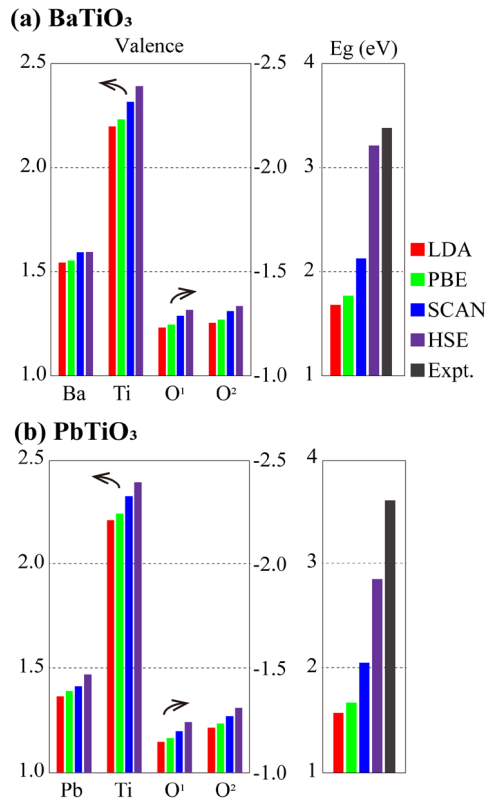

**Figure 1. Valence states and bandgaps of (a) BaTiO$_3$ and (b) PbTiO$_3$.** The valence refers to the difference between the calculated electron occupation and that of neutral atoms. Oxygen atoms are categorized into O$^1$ and O$^2$ to reflect their different chemical environments in the ferroelectric phase. The experimental bandgaps are 3.27 ~ 3.38 eV for BaTiO$_3$ and 3.6 eV for PbTiO$_3$ [25].

The improvement with SCAN lies fundamentally in a better description of the anisotropic bonding and electron distribution around the Ti atoms. It is well-known that Ti's off-center displacement is driven by the covalent bonding between O-2$p$ and Ti-3$d$ in a shorter Ti-O bond, competing with a longer Ti-O bond exhibiting more ionic characteristics [8]. The covalency and ionicity of these bonds can be deduced from the atomic valences. Figure 1 compares the valences calculated using LDA, PBE, SCAN, and HSE06. Overall, the ionicity in BaTiO$_3$ and PbTiO$_3$ monotonically increases as the exchange-correlation ladder progresses, i.e., from LDA to PBE to SCAN and finally to the HSE06 hybrid functional.

The enhanced ionicity of the Ti-O bond from PBE to SCAN contributes to increased bandgaps (Figure 1), which are 1.75 eV from PBE and 2.1 eV from SCAN for BaTiO$_3$. The valence band of BaTiO$_3$ is primarily composed of Ti-3$d$ and O-2$p$ electrons, as evidenced by the fact that Ti-3$d$ orbitals contain 1.97 electrons in the PBE calculation. Significant $p$-$d$ Coulombic repulsion pushes electrons toward the band edge, which sensitively determines the bandgap size. The enhanced ionicity indicates inter-site electron redistribution from Ti to O, as demonstrated by the reduction of Ti-3$d$ electrons to 1.89 in the SCAN calculation. Accordingly, from PBE to SCAN, 0.08 electrons are transferred from Ti to O. Such electron redistribution (from PBE to SCAN) reduces the $p$-$d$ repulsion, lowering the valence band positions and thus increasing the bandgap. The improvement in BaTiO$_3$ is quantitatively limited, and we will later reveal a more profound consequence in YMnO$_3$.

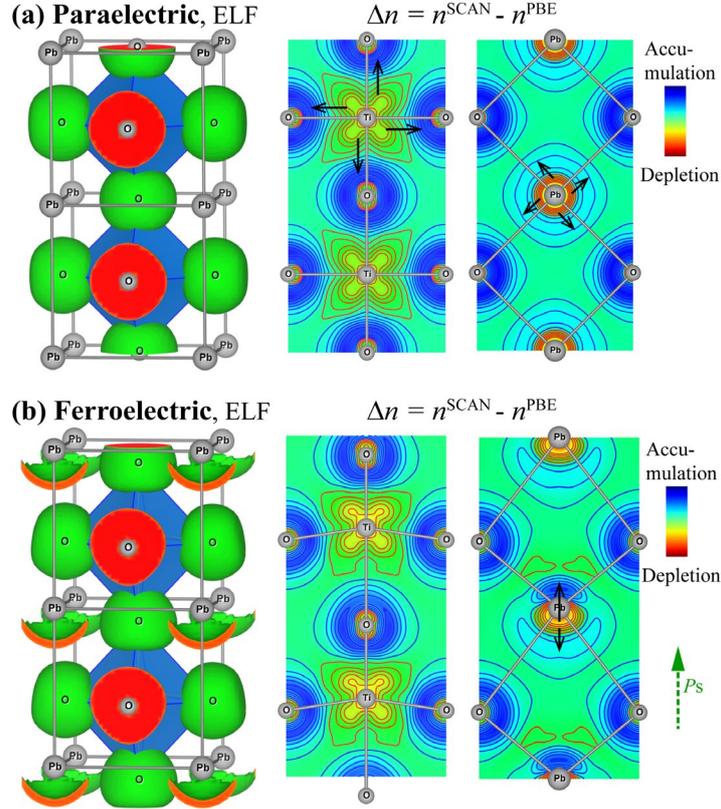

**Figure 2. Electron density analysis in PbTiO$_3$.** (a) The paraelectric phase. The left panel illustrates the electron localization function (ELF) with the green isosurfaces. The middle and right panels display the density changes on two (100) planes. Blue and red colors signify electron accumulation and depletion, respectively. The black arrows highlight the electron redistribution. (b) The ferroelectric phase. The left panel highlights the Pb-6$s^2$ lone-pair electrons. The middle and right panels highlight the anisotropic electron distributions around Ti and Pb atoms.

The above arguments can be visually illustrated through electron density analysis. Figure 2 plots the electron localization function (ELF) and electron redistribution pattern for PbTiO$_3$, encompassing its paraelectric and ferroelectric phases. In the paraelectric phase (Figure 2a), the ELF highlights electron accumulation around the oxygen anions, while the Pb and Ti cations retain fewer electrons. From PBE to SCAN calculations, the valence electrons undergo significant redistribution, as observed from $\Delta n = n^{\text{SCAN}} - n^{\text{PBE}}$. On one hand, the Ti atoms lose some electrons and become more positive, while the oxygen atoms attract more electrons and become more negative. On the other hand, oxygen also pulls electrons away from Pb sites, leading to electron depletion at the site center and accumulation around the site border. These two trends collectively enhance the ionic nature of the bonds in PbTiO$_3$ within the SCAN simulations [25]. This increase in Ti-O ionicity, which consequently reduces covalency, weakens the Ti's polar displacement, as the ferroelectric distortion is primarily driven by $pd$ covalent bonding [8]. Therefore, the *super-tetragonality* error is alleviated by SCAN.

When polarized (Figure 2b), Pb-6$s^2$ lone-pair electrons become stereochemically active, as evidenced by the cap-shaped ELF below the Pb ions. The Pb atom undergoes non-centrosymmetric displacement alongside Ti, and this double-driven force results in PbTiO$_3$ exhibiting much stronger ferroelectric polarization than BaTiO$_3$ [2]. The electron redistribution, $\Delta n$, reveals an interesting pattern around Pb sites: electrons transfer from the lower side to the upper side of Pb, which counters the cap-shaped electron distribution and thus further mitigates the *super-tetragonality* problem.

### 3.2. Localization of open *d*-shells

Multiferroic materials of BiFeO$_3$ [34,35] and YMnO$_3$ [36] introduce an extra complexity due to their open 3*d* shells. Since both materials are semiconductors, establishing finite bandgaps is usually essential for reliable DFT calculations. Figure 3 shows the calculation results: LDA mistakenly identifies BiFeO$_3$ as a metal, while no functional predicts a bandgap for YMnO$_3$. This computational challenge is often attributed to an inadequate description of electron localization. A simple remedy is to augment DFT with a Hubbard-like *U* parameter [37,38], where the localized orbitals receive separate treatment. In this approach, the localized electrons are subject to a corrective potential [37], $V_{\text{DFT}+U} = V_{\text{DFT}} + U(\frac{1}{2} - n_i)$, where $n_i = 1$ for a filled orbital and zero for an empty one. An electron in a filled orbital is subject to a potential reduced by *U*/2, which lowers its energy and increases its spatial localization [17,38]. For BiFeO$_3$, a modest *U* value within LDA+*U* can rectify the bandgap prediction. Even higher *U* values tend to saturate the bandgap, indicating that the Fe-3*d* states are energetically too distant from the band edge to notably influence the bandgap.

For BiFeO$_3$, SCAN yields quantitatively similar magnetic moments and bandgaps (see Figure 3a) to PBE+*U* with *U* = 2 eV, hinting at some similar effects of employing +*U* correction and the SCAN functional [16]. To gain insights, we explore the electron densities. Starting with the PBE calculation, density modifications are made using the *U* correction (i.e., $\Delta n^{\text{a}} = n^{\text{PBE}+\text{U}} - n^{\text{PBE}}$) and SCAN functional (i.e., $\Delta n^{\text{b}} = n^{\text{SCAN}} - n^{\text{PBE}}$), as plotted in Figure 4a. First, the overall similarity between $\Delta n^{\text{a}}$ and $\Delta n^{\text{b}}$ provides clear evidence of SCAN's ability to localize *d* orbitals, as achieved by the *U* correction. For example, SCAN and PBE+*U* yield almost identical magnetic moments of ~3.9 $\mu_\text{B}$ for Fe in BiFeO$_3$ (Figure 3a). Quantitatively, SCAN can correct the delocalization error in PBE by approximately 2 eV. Second, $\Delta n^{\text{a}}$ and $\Delta n^{\text{b}}$ patterns clearly reveal substantial electron depletion in the Fe-$t_{2g}$ orbitals, but a relatively weak electron accumulation in the $e_\text{g}$ orbitals. It indicates that the *U* values should be orbital-dependent to account for the different orbital environments, which makes the DFT+*U* method technically involved. By contrast, an advantage of SCAN is its inherent capability to recognize chemical bonds and environments [14], including the anisotropic bonding behaviors of the *d* orbitals. Besides the orbital localization, SCAN's electron modulation reveals effective electron transfer from Fe to O, and an electron redistribution on Bi sites. As mentioned earlier, such electron modulation has a remarkable impact on the bandgap: SCAN predicts a more realistic bandgap of 1.8 eV compared with PBE's result of 0.8 eV.

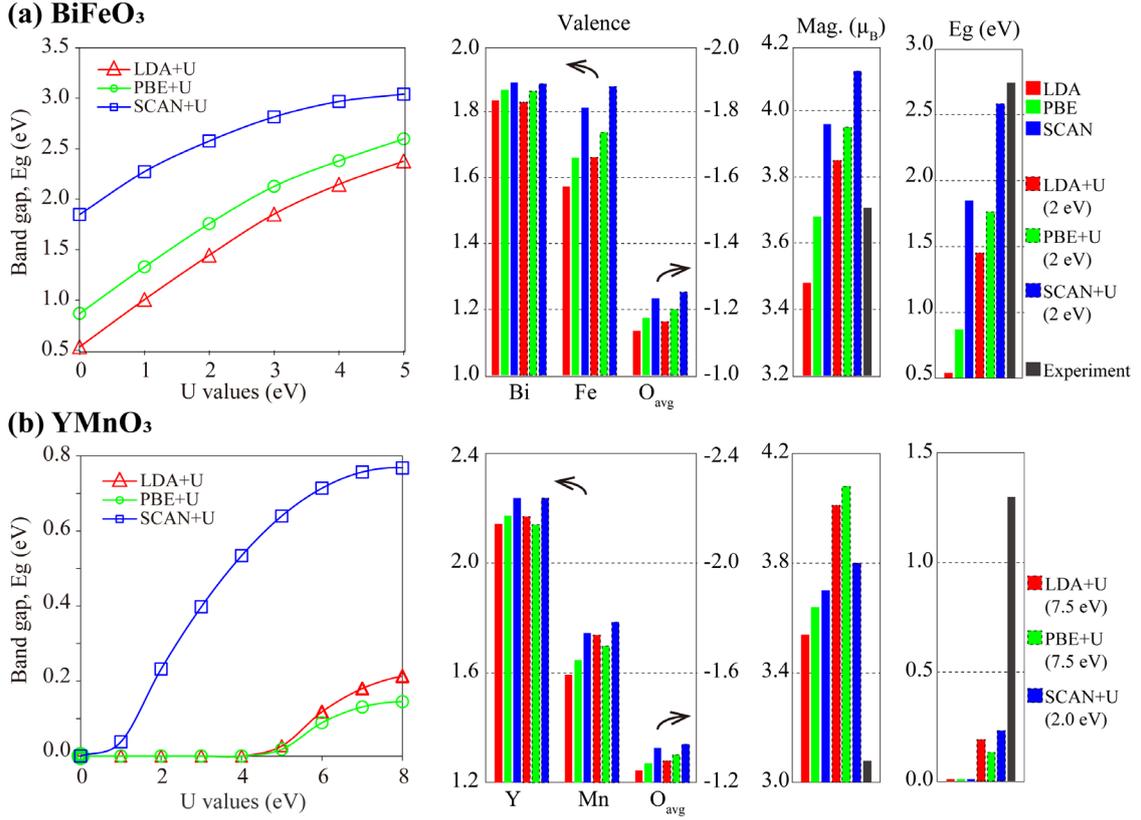

**Figure 3. Electronic properties of (a) BiFeO$_3$ and (b) YMnO$_3$.** The first column displays the bandgaps for varying $U$ values across the LDA+$U$, PBE+$U$, and SCAN+$U$ methods. The second panel shows ionic valences, with oxygen valences averaged when they exist in different chemical environments. The third and fourth panels detail the local magnetic moments and the bandgaps, respectively. Experimental geometries are employed to ensure consistent comparisons. The experimental bandgaps are 2.67 eV [39] for BiFeO$_3$ and 1.35 eV for YMnO$_3$ [40].

YMnO$_3$ has an experimental bandgap of 1.35 eV [40], but no functional can open a bandgap, as illustrated in Figure 3b. The minimum $U$ required to induce a bandgap varies significantly: it is as high as 5.0 eV for PBE+$U$ but only 1.0 eV for SCAN+$U$. If $d$-electron delocalization were the only problem in the PBE calculation, it would be perplexing that SCAN can reduce the delocalization error by as much as 4 eV. To resolve this puzzle, Figure 4b compares the differential electron densities, $\Delta n^c = n^{PBE+U} - n^{PBE}$ and $\Delta n^d = n^{SCAN+U} - n^{PBE}$. $\Delta n^c$ predominantly shows electron accumulation in the Mn-$d_z^2$ orbitals, a trend also observed in $\Delta n^d$. However, $\Delta n^d$ exhibits additional characteristics: significant electron transfer from Mn-$d_z^2$ orbitals to oxygen sites, and substantial electron accumulation in the planar $d_{xy}$ and $d_{x^2-y^2}$ orbitals. This inter-site electron transfer enables SCAN to open a bandgap with a much smaller $U$, as detailed in the next section. Moreover, this orbital-dependent behavior is evidence of SCAN's robust capability to capture complex anisotropic bonding environments.

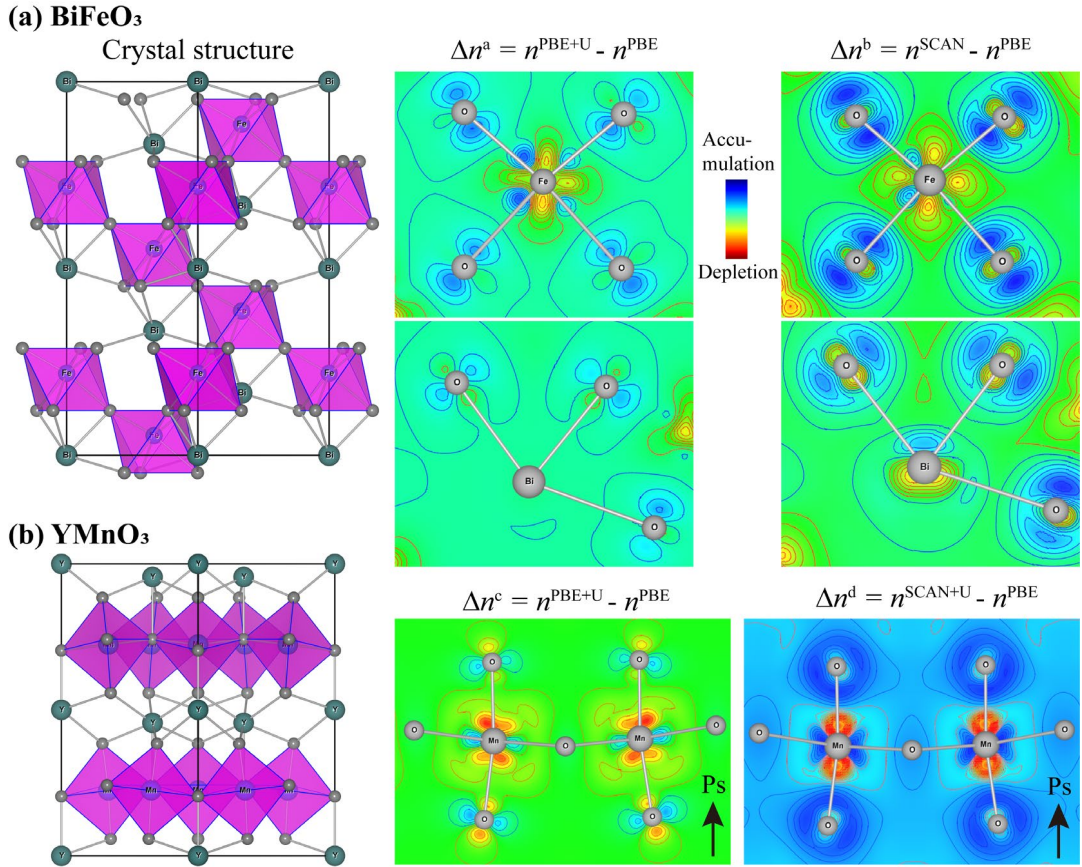

**Figure 4. Electron density analysis in (a) BiFeO$_3$ and (b) YMnO$_3$.** The crystal structures are presented in the left panel. The middle and right panels illustrate the differential electron distributions, $\Delta n = n^{PBE+U} - n^{PBE}$ and $\Delta n = n^{SCAN} - n^{PBE}$. Electron depletion (i.e., $\Delta n < 0$, shown in reddish colors) occurs because the orbitals become more localized or some electrons are transferred away from this region. An identical density range is employed throughout the plots to enable quantitative comparisons. For the PBE+$U$ calculations, the $U$ values are 2.0 and 7.5 eV for BiFeO$_3$ and YMnO$_3$, respectively. The $U$ value is 2 eV in SCAN+$U$ for YMnO$_3$.

### 3.3. Synergistic effects on the band structure calculation

Compared to the PBE+$U$ method, the SCAN approach not only similarly localizes the $d$-orbitals but also uniquely enhances the accuracy in describing ionicity and orbital anisotropy. It raises a pertinent question: How do these enhancements synergistically allow for a much smaller $U$ value in SCAN+$U$ for YMnO$_3$? To answer this, it is crucial to examine the differences in electronic properties between BiFeO$_3$ and YMnO$_3$, as depicted in Figures 5 and 6.

For BiFeO$_3$, the upper valence bands in the PBE simulation exhibit considerable $p$-$d$ hybridization (Figure 5a). However, Fe-3$d$ states at the band edge (0 to −0.5 eV) are excessively pronounced compared to experimental results, suggesting a problem of $d$-orbital under-binding in energy (or delocalization in the real space) by PBE. PBE+$U$ (2 eV) effectively lowers the Fe-3$d$ orbitals in the valence bands and pushes the conduction bands upward. While the valence band structure is extensively altered due to pronounced $p$-$d$ hybridization, the isolated conduction bands undergo a rigid shift (Figure 6a). The experimental X-ray photoelectron spectrum (for instance, peaks A and C) is well reproduced by the PBE+$U$ simulation. SCAN's spectrum resembles that of PBE+$U$, but a particularly notable outcome from SCAN is peak B's emergence, making SCAN's results more reliable. SCAN+$U$ (2 eV) and HSE lead to over-corrections: the Fe-3$d$ orbitals become too deep, markedly diminishing $p$-$d$ hybridization and resulting in overly narrow valence bands. For YMnO$_3$, both PBE and SCAN calculations incorrectly predict metallic behavior. Figure 6b also shows that the Mn-4$s$ states, which are supposed to be in the conduction bands, are incorrectly inverted into the valence band near the Brillouin Γ point. This problematic topology is corrected when a bandgap is introduced using PBE+$U$ with $U$ = 7.5 eV and SCAN+$U$ with $U$ = 2 eV. However, $U$ = 7.5 eV is so large that it excessively displaces Mn-3$d$ states away from the band edge, resulting in a significant discrepancy with experimental spectroscopy, as shown in Figure 5b. The valence band spectroscopy and orbital ordering are relatively well calculated by SCAN+$U$ ($U$ = 2 eV).

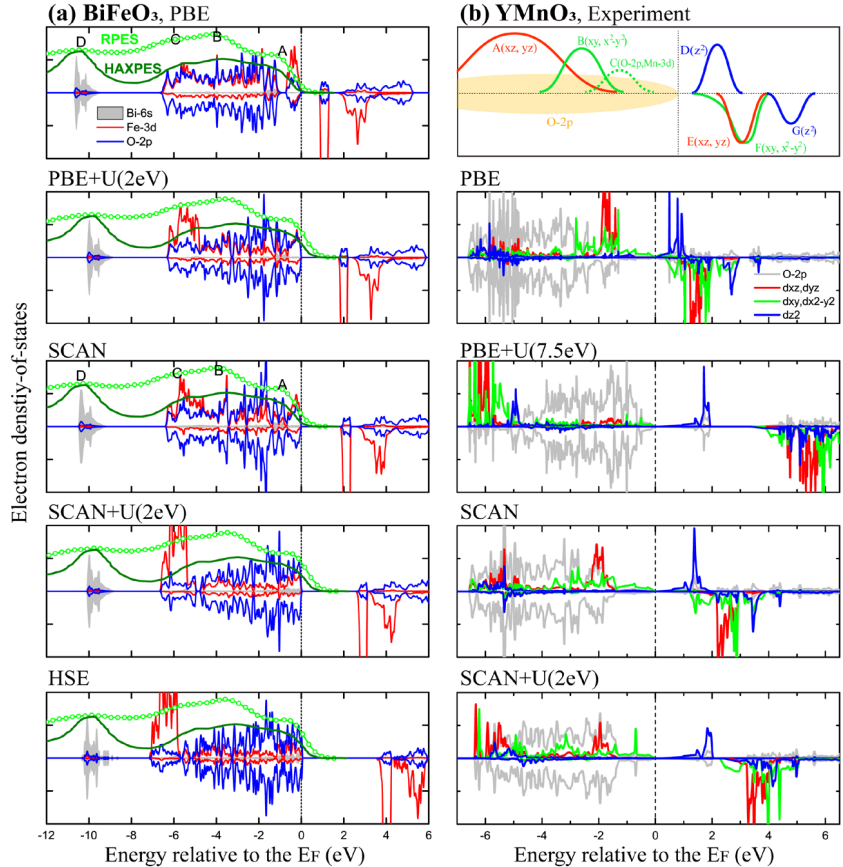

**Figure 5. The density-of-states for (a) BiFeO$_3$ and (b) YMnO$_3$.** BiFeO$_3$: Hard X-ray photoelectron spectroscopy (HAXPES) harnesses a photon energy of 2010 eV to probe the collective valence states, emphasizing the contributions from Fe-3$d$ and O-2$p$ [41]. Resonant photoelectron spectroscopy (RPES) predominantly captures the Fe-3$d$ states [41]. The calculated and experimental profiles are aligned at the Bi-6$s^2$ lone-pair states around −10 eV, labeled as peak D. For clarity in comparison, the experimental spectra are arbitrarily vertically displaced. YMnO$_3$: The illustrative valence band diagram is derived from resonance photoemission [42], while the conduction band diagram is informed by correlating photoemission spectroscopy (PES) with X-ray absorption spectroscopy (XAS) [43]. Peak C primarily contains O-2$p$ states, with a notable mixture of Mn-3$d$ states.

Compared with BiFeO$_3$, a dramatic difference in YMnO$_3$ is that both the near-edge bands are highly dispersive. Figure 6(b) reveals that the top valence bands are derived from planar Mn-3$d$ orbitals (i.e., $dxy$ and $dx^2$-$y^2$) and O-2$p$ orbitals, and the directional orbitals have a strong σ-type overlap. SCAN not only makes the Mn-3$d$ orbitals more localized, but also captures more ionicity in the Mn-O bond (refer to Figure 4). This observation holds true for the bottom conduction bands derived from σ-type overlap between Mn-4$s$ and Mn-d$z^2$ orbitals. SCAN better calculates the Mn-4$s$ orbitals, an unattainable feature by the Hubbard-like $U$ method. As a result, the simultaneous improvement of the $s$, $p$, and $d$ orbitals in SCAN enables it to open a bandgap without significantly distorting the overall band structure. Nevertheless, a small Hubbard-like $U$ is still needed to address the remaining delocalization error in SCAN.

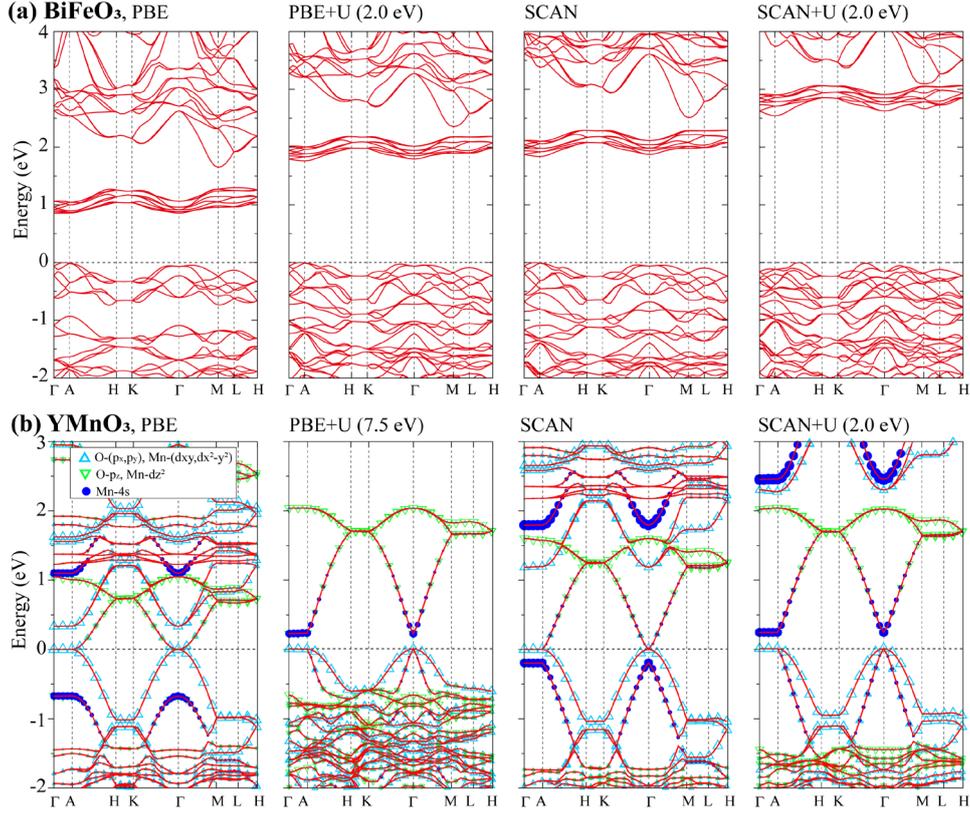

**Figure 6. Band structures of (a) BiFeO$_3$ and (b) YMnO$_3$.** For YMnO$_3$, wave functions are projected onto various orbitals. In the PBE+$U$ calculation with $U$ = 7.5 eV, the conduction bands derived from the planar Mn $d$ orbitals (i.e., $dxy$ and $dx^2$-$y^2$) lie outside the energy window due to the excessively high $U$ value. All calculations were performed using relaxed geometries corresponding to each respective method.

## 4. DISCUSSION AND SUMMARY

The local magnetic moments computed using the SCAN functional, $M_{Fe}$ = 3.96 $\mu_B$ in BiFeO$_3$ and $M_{Mn}$ = 3.70 $\mu_B$ in YMnO$_3$, exceed the experimental values of 3.75 $\mu_B$ [44] and 3.09 $\mu_B$ [45], respectively. Such an overestimation tendency was first noted in elemental iron [46]. However, it is crucial to recognize that experimental values can be influenced by factors not typically accounted for by DFT. For example, magnetic moments can decrease due to thermal effects and spin fluctuations [47,48]. In the specific case of YMnO$_3$, the triangular lattice of Mn ions introduces geometrical frustration, which could significantly amplify spin fluctuations. It is worth noting that the experimental value of $M_{Mn}$ = 3.09 $\mu_B$ [45] is much lower than all the theoretical predictions, including the LDA method. However, magnetization overestimation by LDA is unusual because LDA typically has a high delocalization error [15]. This theoretical-experimental discrepancy highlights the need for further research, emphasizing thermal and quantum effects not inherently included in standard DFT simulations.

This study demonstrates that the SCAN meta-GGA functional consistently outperforms traditional functionals in calculating the fundamental electronic properties of complex ferroelectric and multiferroic materials. SCAN improvements over the widely-used PBE functional are evident through several computational issues: the super-tetragonality of BaTiO$_3$ and PbTiO$_3$, and the fidelity of band structures of BiFeO$_3$ and YMnO$_3$. The enhanced performance of SCAN is attributed to its reduction of delocalization error, which is manifested in three characteristics: more compact orbitals, more accurate ionicity, and better orbital anisotropy. In comparison, the primary and straightforward consequence of applying a Hubbard-like U correction is the increased localization of the $d$-orbitals. SCAN's advancements hold significant implications for the computational materials science community, and investigating electron density also provides valuable insights to address density-driven errors.

# ACKNOWLEDGEMENT

YZ is supported by the Natural Science Foundation of Fujian Province (2023J02032) and the National Natural Science Foundation of China (11904156). DK is supported by the Natural Science Foundation of Fujian Province (2022J011127). JS acknowledges the support by the U.S. Office of Naval Research (ONR) Grant No. N00014-22-1-2673.

# REFERENCES

[1] Andrew L. Goodwin, *Opportunities and challenges in understanding complex functional materials.* Nature Communications. 10, 4461 (2019).
[2] Karin M Rabe, Charles H Ahn, Jean-Marc Triscone, *Physics of ferroelectrics: a modern perspective*. Vol. 105. 2007: Springer Science & Business Media.
[3] Manfred Fiebig, Thomas Lottermoser, Dennis Meier, Morgan Trassin, *The evolution of multiferroics.* Nature Reviews Materials. 1, 16046 (2016).
[4] G. Jeffrey Snyder, Eric S. Toberer, *Complex thermoelectric materials.* Nat. Mater. 7, 105 (2008).
[5] B. Keimer, S. A. Kivelson, M. R. Norman, S. Uchida, J. Zaanen, *From quantum matter to high-temperature superconductivity in copper oxides.* Nature. 518, 179 (2015).
[6] Walter Kohn, Lu Jeu Sham, *Self-consistent equations including exchange and correlation effects.* Phys. Rev. 140, A1133 (1965).
[7] Pierre Hohenberg, Walter Kohn, *Inhomogeneous electron gas.* Phys. Rev. 136, B864 (1964).
[8] Ronald E Cohen, *Origin of ferroelectricity in perovskite oxides.* Nature. 358, 136 (1992).
[9] John P. Perdew, *Density-functional approximation for the correlation energy of the inhomogeneous electron gas.* Phys. Rev. B. 33, 8822 (1986).
[10] John P. Perdew, Yue Wang, *Accurate and simple analytic representation of the electron-gas correlation energy.* Phys. Rev. B. 45, 13244 (1992).
[11] John P Perdew, Kieron Burke, Matthias Ernzerhof, *Generalized gradient approximation made simple.* Phys. Rev. Lett. 77, 3865 (1996).
[12] Jianwei Sun, Adrienn Ruzsinszky, John P Perdew, *Strongly constrained and appropriately normed semilocal density functional.* Phys. Rev. Lett. 115, 036402 (2015).
[13] Ryan Kingsbury, Ayush S. Gupta, Christopher J. Bartel, Jason M. Munro, Shyam Dwaraknath, Matthew Horton, Kristin A. Persson, *Performance comparison of r2SCAN and SCAN metaGGA density functionals for solid materials via an automated, high-throughput computational workflow.* Physical Review Materials. 6, 013801 (2022).
[14] Jianwei Sun, Bing Xiao, Yuan Fang, Robin Haunschild, Pan Hao, Adrienn Ruzsinszky, Gábor I Csonka, Gustavo E Scuseria, John P Perdew, *Density functionals that recognize covalent, metallic, and weak bonds.* Phys. Rev. Lett. 111, 106401 (2013).
[15] J. P. Perdew, Alex Zunger, *Self-interaction correction to density-functional approximations for many-electron systems.* Phys. Rev. B. 23, 5048 (1981).
[16] Yubo Zhang, James W. Furness, Bing Xiao, Jianwei Sun, *Subtlety of TiO2 phase stability: Reliability of the density functional theory predictions and persistence of the self-interaction error.* J. Chem. Phys. 150, (2019).
[17] Yubo Zhang, James Furness, Ruiqi Zhang, Zhi Wang, Alex Zunger, Jianwei Sun, *Symmetry-breaking polymorphous descriptions for correlated materials without interelectronic U.* Phys. Rev. B. 102, 045112 (2020).
[18] Alex Zunger, *Bridging the gap between density functional theory and quantum materials.* Nature Computational Science. 2, 529 (2022).
[19] Julien Varignon, Manuel Bibes, Alex Zunger, *Origin of band gaps in 3d perovskite oxides.* Nature Communications. 10, 1658 (2019).
[20] Yubo Zhang, Da Ke, Junxiong Wu, Chutong Zhang, Lin Hou, Baichen Lin, Zuhuang Chen, John P. Perdew, Jianwei Sun, *Challenges for density functional theory in simulating metal–metal singlet bonding: A case study of dimerized VO2.* J. Chem. Phys. 160, (2024).
[21] James W. Furness, Yubo Zhang, Christopher Lane, Ioana Gianina Buda, Bernardo Barbiellini, Robert S. Markiewicz, Arun Bansil, Jianwei Sun, *An accurate first-principles treatment of doping-dependent electronic structure of high-temperature cuprate superconductors.* Communications Physics. 1, 11 (2018).
[22] Yubo Zhang, Christopher Lane, James W. Furness, Bernardo Barbiellini, John P. Perdew, Robert S. Markiewicz, Arun Bansil, Jianwei Sun, *Competing stripe and magnetic phases in the cuprates from first principles.* Proceedings of the National Academy of Sciences. 117, 68 (2020).

[23] Christopher Lane, James W. Furness, Ioana Gianina Buda, Yubo Zhang, Robert S. Markiewicz, Bernardo Barbiellini, Jianwei Sun, Arun Bansil, *Antiferromagnetic ground state of La$_2$CuO$_4$: A parameter-free ab initio description.* Phys. Rev. B. 98, 125140 (2018).

[24] Ruiqi Zhang, Bahadur Singh, Christopher Lane, Jamin Kidd, Yubo Zhang, Bernardo Barbiellini, Robert S. Markiewicz, Arun Bansil, Jianwei Sun, *Critical role of magnetic moments in heavy-fermion materials: Revisiting SmB6.* Phys. Rev. B. 105, 195134 (2022).

[25] Yubo Zhang, Jianwei Sun, John P. Perdew, Xifan Wu, *Comparative first-principles studies of prototypical ferroelectric materials by LDA, GGA, and SCAN meta-GGA.* Phys. Rev. B. 96, 035143 (2017).

[26] Min-Cheol Kim, Eunji Sim, Kieron Burke, *Understanding and Reducing Errors in Density Functional Calculations.* Phys. Rev. Lett. 111, 073003 (2013).

[27] Georg Kresse, Jürgen Furthmüller, *Efficient iterative schemes for ab initio total-energy calculations using a plane-wave basis set.* Phys. Rev. B. 54, 11169 (1996).

[28] Peter E Blöchl, *Projector augmented-wave method.* Phys. Rev. B. 50, 17953 (1994).

[29] Georg Kresse, D Joubert, *From ultrasoft pseudopotentials to the projector augmented-wave method.* Phys. Rev. B. 59, 1758 (1999).

[30] Jochen Heyd, Gustavo E Scuseria, Matthias Ernzerhof, *Hybrid functionals based on a screened Coulomb potential.* J. Chem. Phys. 118, 8207 (2003).

[31] Joachim Paier, Martijn Marsman, K Hummer, Georg Kresse, Iann C Gerber, János G Ángyán, *Screened hybrid density functionals applied to solids.* J. Chem. Phys. 124, 154709 (2006).

[32] Graeme Henkelman, Andri Arnaldsson, Hannes Jónsson, *A fast and robust algorithm for Bader decomposition of charge density.* Computational Materials Science. 36, 354 (2006).

[33] DI Bilc, R Orlando, Riad Shaltaf, G-M Rignanese, Jorge Íñiguez, Ph Ghosez, *Hybrid exchange-correlation functional for accurate prediction of the electronic and structural properties of ferroelectric oxides.* Phys. Rev. B. 77, 165107 (2008).

[34] J. Wang, J. B. Neaton, H. Zheng, V. Nagarajan, S. B. Ogale, B. Liu, D. Viehland, V. Vaithyanathan, D. G. Schlom, U. V. Waghmare, N. A. Spaldin, K. M. Rabe, M. Wuttig, R. Ramesh, *Epitaxial BiFeO3 Multiferroic Thin Film Heterostructures.* Science. 299, 1719 (2003).

[35] J. B. Neaton, C. Ederer, U. V. Waghmare, N. A. Spaldin, K. M. Rabe, *First-principles study of spontaneous polarization in multiferroic BiFeO$_3$.* Phys. Rev. B. 71, 014113 (2005).

[36] Bas B Van Aken, Thomas TM Palstra, Alessio Filippetti, Nicola A Spaldin, *The origin of ferroelectricity in magnetoelectric YMnO3.* Nat. Mater. 3, 164 (2004).

[37] I. Anisimov Vladimir, F. Aryasetiawan, A. I. Lichtenstein, *First-principles calculations of the electronic structure and spectra of strongly correlated systems: the LDA+ U method.* J. Phys.: Condens. Matter. 9, 767 (1997).

[38] Matteo Cococcioni, Stefano de Gironcoli, *Linear response approach to the calculation of the effective interaction parameters in the LDA+U method.* Phys. Rev. B. 71, 035105 (2005).

[39] Daniel Sando, Cécile Carrétéro, Mathieu N. Grisolia, Agnès Barthélémy, Valanoor Nagarajan, Manuel Bibes, *Revisiting the Optical Band Gap in Epitaxial BiFeO3 Thin Films.* Advanced Optical Materials. 6, 1700836 (2018).

[40] Lei Chen, Guifang Zheng, Gang Yao, Pingjuan Zhang, Shangkai Dai, Yang Jiang, Heqin Li, Binbin Yu, Haiyong Ni, Shizhong Wei, *Lead-Free Perovskite Narrow-Bandgap Oxide Semiconductors of Rare-Earth Manganates.* ACS Omega. 5, 8766 (2020).

[41] Dipanjan Mazumdar, R Knut, F Thöle, M Gorgoi, Sergei Faleev, ON Mryasov, Vilas Shelke, C Ederer, NA Spaldin, A Gupta, *The valence band electronic structure of rhombohedral-like and tetragonal-like BiFeO3 thin films from hard X-ray photoelectron spectroscopy and first-principles theory.* J. Electron Spectrosc. Relat. Phenom. 208, 63 (2015).

[42] Manish Kumar, RJ Choudhary, DM Phase, *Valence band structure of YMnO3 and the spin orbit coupling.* Appl. Phys. Lett. 102, 182902 (2013).

[43] J-S Kang, SW Han, J-G Park, SC Wi, SS Lee, G Kim, HJ Song, HJ Shin, W Jo, BI Min, *Photoemission and x-ray absorption of the electronic structure of multiferroic RMnO3 (R= Y, Er).* Phys. Rev. B. 71, 092405 (2005).

[44] I Sosnowska, W Schäfer, W Kockelmann, KH Andersen, IO Troyanchuk, *Crystal structure and spiral magnetic ordering of BiFeO3 doped with manganese.* Applied Physics A. 74, s1040 (2002).

[45] M Janoschek, B Roessli, L Keller, SN Gvasaliya, K Conder, E Pomjakushina, *Reduction of the ordered magnetic moment in YMnO3 with hydrostatic pressure.* J. Phys.: Condens. Matter. 17, L425 (2005).

[46] Yuhao Fu, David J. Singh, *Applicability of the Strongly Constrained and Appropriately Normed Density Functional to Transition-Metal Magnetism.* Phys. Rev. Lett. 121, 207201 (2018).

[47] L Ortenzi, II Mazin, P Blaha, L Boeri, *Accounting for spin fluctuations beyond local spin density approximation in the density functional theory.* Phys. Rev. B. 86, 064437 (2012).

[48] Anna Galler, Ciro Taranto, Markus Wallerberger, Merzuk Kaltak, Georg Kresse, Giorgio Sangiovanni, Alessandro Toschi, Karsten Held, *Screened moments and absence of ferromagnetism in FeAl.* Phys. Rev. B. 92, 205132 (2015).